\documentclass[12pt]{article}

\usepackage{axodraw}
\usepackage{epsf}
\usepackage{rotating}
\usepackage{color}
\newcommand{\beq}{\begin{equation}}
\newcommand{\eeq}{\end{equation}}
\newcommand{\bea}{\begin{eqnarray}}
\newcommand{\eea}{\end{eqnarray}}
\newcommand{\bed}{\begin{displaymath}}
\newcommand{\eed}{\end{displaymath}}

\newcommand{\tga}{{\rm tg}\alpha}
\newcommand{\tgb}{{\rm tg}\beta}
\newcommand{\stgb}{{\rm tg}^2\beta}
\newcommand{\sgl}{\tilde{g}}

\newcommand{\st}{\tilde t}
\newcommand{\ssb}{\tilde b}
\newcommand{\sq}{\tilde q}

\newcommand{\gl}{\tilde g}

\newcommand{\gev}{~\mbox{GeV}}

\newcommand{\lsim}{\raisebox{-0.13cm}{~\shortstack{$<$ \\[-0.07cm] $\sim$}}~}

\catcode`\@=11

\def\@citex[#1]#2{\if@filesw\immediate\write\@auxout{\string\citation{#2}}\fi
  \def\@citea{}\@cite{\@for\@citeb:=#2\do
    {\@citea\def\@citea{,\penalty\@m}\@ifundefined
       {b@\@citeb}{{\bf ?}\@warning
       {Citation `\@citeb' on page \thepage \space undefined}}%
\hbox{\csname b@\@citeb\endcsname}}}{#1}}

\def\citer{\@ifnextchar [{\@tempswatrue\@citexr}{\@tempswafalse\@citexr[]}}
\def\@citexr[#1]#2{\if@filesw\immediate\write\@auxout{\string\citation{#2}}\fi
  \def\@citea{}\@cite{\@for\@citeb:=#2\do
    {\@citea\def\@citea{--\penalty\@m}\@ifundefined
       {b@\@citeb}{{\bf ?}\@warning
       {Citation `\@citeb' on page \thepage \space undefined}}%
\hbox{\csname b@\@citeb\endcsname}}}{#1}}
\catcode`\@=12

\begin{document}

\renewcommand{\thefootnote}{\fnsymbol{footnote}}
\setcounter{page}{0}

\begin{titlepage}

\vskip-1.0cm

\vspace*{2em}

\begin{center}
{\large\bf  Neutral MSSM Higgs-Boson Production With Heavy Quarks:\\[0.25cm]
NLO Supersymmetric QCD Corrections} \\
\end{center}

\vskip 1.cm
\begin{center}
{\sc S.~Dittmaier$^1$, P.~H\"afliger$^{2,3}$, M.~Kr\"amer$^4$,
M.~Spira$^3$ and M.~Walser$^{3,5}$}

\vskip 0.8cm

\begin{small} 
{\it \small
$^1$ Physikalisches Institut, Albert--Ludwigs--Universit\"at Freiburg,
D--79104 Freiburg, Germany \\
$^2$ Institute for Particle Physics, ETH Z\"urich, CH--8093 Z\"urich,
Switzerland \\
$^3$ Paul Scherrer Institut, CH--5232 Villigen PSI, Switzerland \\
$^4$ Institute for Theoretical Particle Physics and Cosmology, RWTH Aachen University, D-52056 Aachen,
Germany\\
$^5$ Institute for Theoretical Physics, ETH Z\"urich, CH--8093 Z\"urich,
Switzerland
}
\end{small}
\end{center}

\vskip 2cm

\begin{abstract}
\noindent
Within the minimal supersymmetric extension of the Standard Model (MSSM)
the associated production of neutral Higgs bosons with top and
bottom quarks belongs to the most important Higgs-boson production
processes at the LHC. At large values of $\tgb$, in particular, bottom--Higgs 
associated production constitutes the dominant production channel within 
the MSSM. We have calculated the next-to-leading-order
supersymmetric QCD corrections to neutral Higgs production through the
parton processes $q\bar q, gg \to t\bar t/b\bar b + h/H/A$ and present
results for the total cross sections. The genuine SUSY-QCD corrections
are of moderate size for small $\tgb$, but can be sizable for large $\tgb$.
In the latter case the bulk of these 
corrections can be absorbed into effective bottom Yukawa
couplings.
\end{abstract}

\vfill
June 2014

\end{titlepage}

\renewcommand{\thefootnote}{\arabic{footnote}}

\setcounter{footnote}{0}

\section{Introduction}
The Higgs mechanism~\cite{hi64} is a cornerstone of the Standard Model
(SM) and its supersymmetric extensions.  
Even after the discovery~\cite{higgsdiscovery}
of a Higgs boson by the LHC experiments ATLAS and CMS
in 2012, Higgs-boson searches
belong to the major endeavors at present and future colliders, 
but now with a particular focus on extended Higgs sectors.
The minimal supersymmetric extension of the SM (MSSM) is among the
best motivated models providing more than one Higgs boson.
In the
MSSM two isospin Higgs doublets are introduced in order to generate
masses for up- and down-type fermions \cite{twoiso}. After electroweak
symmetry breaking three of the eight degrees of freedom are absorbed by
the $Z$ and $W$ gauge bosons, leaving five states as elementary Higgs
particles. These consist of two CP-even neutral (scalar) particles
$h,H$, one CP-odd neutral (pseudoscalar) particle $A$, and two charged
bosons $H^\pm$.  At leading order the MSSM Higgs sector is fixed by two
independent input parameters which are usually chosen as the
pseudoscalar Higgs mass $M_A$ and $\tgb=v_2/v_1$, the ratio of the two
vacuum expectation values. 
 Including the one-loop and dominant two-loop
corrections the upper bound of the light scalar Higgs mass is $M_h\lsim
135$~GeV \cite{mssmrad} for supersymmetric mass scales up to about a TeV.
Three-loop results \cite{mssmrad3} and the resummation of logarithmic corrections 
  from the scalar top sector \cite{Barbieri:1990ja} confirm this upper
bound within 1~GeV, see the recent numerical analysis \cite{Hahn:2013ria}.
An important property of the bottom
Yukawa couplings is their enhancement for large values of $\tgb$. 
The top Yukawa couplings, on the other hand, are suppressed for large $\tgb$
\cite{schladming}, unless the light (heavy) scalar Higgs mass is close to its 
upper (lower) bound, where their couplings become SM-like.
The
couplings of the various neutral MSSM Higgs bosons to fermions and gauge
bosons, normalized to the SM Higgs couplings, are listed in
Table~\ref{tb:hcoup}, where the angle $\alpha$ denotes the mixing angle
of the scalar Higgs bosons $h,H$.
\begin{table}[h]
\renewcommand{\arraystretch}{1.5}
\begin{center}
\begin{tabular}{|lc||ccc|} \hline
\multicolumn{2}{|c||}{$\phi$} & $g^\phi_u$ & $g^\phi_d$ &  $g^\phi_V$ \\
\hline \hline
SM~ & $H$ & 1 & 1 & 1 \\ \hline
MSSM~ & $h$ & $\cos\alpha/\sin\beta$ & $-\sin\alpha/\cos\beta$ &
$\sin(\beta-\alpha)$ \\ & $H$ & $\sin\alpha/\sin\beta$ &
$\cos\alpha/\cos\beta$ & $\cos(\beta-\alpha)$ \\
& $A$ & $ 1/\tgb$ & $\tgb$ & 0 \\ \hline
\end{tabular}
\renewcommand{\arraystretch}{1.2}
\caption[]{\label{tb:hcoup} \it MSSM Higgs couplings to 
$u$- and $d$-type fermions
and gauge bosons [$V=W,Z$] relative to the SM couplings.}
\end{center}
\end{table}
The negative direct searches for neutral MSSM Higgs bosons at LEP2 have lead to
lower bounds of $M_{h,H} > 92.8$ GeV and $M_A > 93.4$ GeV~\cite{lep2}. The LEP2 results 
also exclude the range $0.7 < \tgb < 2.0$ in the MSSM, assuming a SUSY
scale $M_{\rm SUSY}=1$~TeV \cite{lep2}. MSSM Higgs-boson searches have continued at 
the $p\bar{p}$ collider Tevatron, see e.g.\ \cite{Aaltonen:2012zh}, 
 and form the central part of the 
current and future physics program of the Large Hadron Collider (LHC) \cite{atlas_cms_tdrs}.
The proton--proton collider LHC has taken data at the centre-of-mass (CM) energies of 7 and 8~TeV, 
and will start operating near its design energy of 13--14~TeV in 2015. 
The present LHC searches at 7 and 8 TeV CM energy have excluded parts of the MSSM
parameter space for large values of $\tgb$ \cite{lhcmssm}. However, the recent
discovery of a resonance with a mass near 125~GeV
\cite{higgsdiscovery} is a clear indication for the existence of a SM or beyond-the-SM Higgs boson. 
While the properties of the new particle, as determined so far, are consistent with those predicted 
within the SM, the bosonic state at 125~GeV can also be 
interpreted as a supersymmetric Higgs boson. 

Neutral MSSM Higgs boson production at  the LHC is dominated 
by gluon fusion, $gg\to h/H/A$, and by the associated production of a Higgs boson with bottom quarks. 
Gluon fusion is most significant at small and moderate $\tgb$. At large values of  $\tgb$, however, bottom--Higgs associated 
production becomes dominant due to the strongly enhanced bottom Yukawa couplings \cite{cxn}.
Higgs-boson radiation off top quarks \cite{tthlo}
(see Fig.~\ref{fig:diags}a)
\begin{displaymath}
q\bar q/gg\to t\bar t + h/H/A
\end{displaymath}
plays a significant role at the LHC for the light scalar Higgs particle
only. The NLO QCD corrections are identical to those for the SM Higgs boson
with modified top and bottom Yukawa couplings, and are thus of moderate
size \cite{tthnlo,tthnlo2}. The SUSY-QCD corrections, which have been computed
in Refs.~\cite{tthnlosusy,tthnlosusy2,tthnlosusy3}, are of moderate
size as well.
First steps towards more realistic descriptions of these processes
have been made in Ref.~\cite{Frederix:2011zi}, where top-quark and
Higgs-boson decays were modelled in LO precision and QCD parton-shower
corrections were matched to the NLO-corrected prediction of the
production process.
The successful experimental study of $t\bar t + h/H/A$ 
with $h/H/A\to b\bar b$ decays critically
depends on the proper control of the huge irreducible background
from direct $t\bar tb\bar b$ and $t\bar t+2jets$ production.
The NLO QCD corrections to these background processes have been
calculated in recent years~\cite{Bredenstein:2009aj},
but further improvements are still
necessary to fully exploit LHC data, such as e.g.\ the inclusion of
QCD corrections to the top-quark decays. Progress in this direction is
ongoing (see, e.g., Refs.~\cite{Dittmaier:2011ti,Dittmaier:2012vm,Heinemeyer:2013tqa}).

For large values of $\tgb$ Higgs-boson radiation off bottom quarks
\cite{tthlo} (see Fig.~\ref{fig:diags}a)
\begin{displaymath}
q\bar q/gg\to b\bar b + h/H/A
\end{displaymath}
constitutes the dominant Higgs-boson production process. The NLO QCD
corrections can be inferred from the analogous calculation involving top
quarks \cite{tthnlo,tthnlo2}. However, they turn out to be numerically enhanced \cite{bbhnlo}. The main
reason is that the integration over small transverse momenta of the
final-state bottom quarks generates sizable logarithmic contributions.
Those logarithms can be resummed by introducing bottom-quark densities in 
the proton~\cite{Barnett:1987jw}
and by applying the standard DGLAP evolution \cite{dglap}. In this so-called 
five-flavour scheme (5FS) the leading-order process is  
\cite{bbh5lo}
\begin{displaymath}
b\bar b\to h/H/A\,,
\end{displaymath}
where the transverse momenta of the incoming bottom quarks, their
masses, and their off-shellness are neglected at LO.  The
NLO~\cite{bbh5nlo} and NNLO \cite{bbh5nnlo} QCD corrections as well as
the SUSY--electroweak corrections \cite{bbh5elw} to these
bottom-initiated processes have been calculated. They are of moderate size, if the
running bottom Yukawa coupling is introduced at the scale of the
corresponding Higgs-boson mass.  The fully exclusive $gg\to b\bar b +
h/H/A$ process, calculated with four active parton flavours in a
four-flavour scheme (4FS), and the 5FS calculation converge at higher perturbative orders, and there is fair 
numerical agreement between the NLO 4FS and NNLO 5FS cross 
section predictions \cite{bbhnlo,bbhscale,Dittmaier:2011ti,Dittmaier:2012vm}.  
In Ref.~\cite{Harlander:2011aa} a scheme has been proposed to match the 4FS and 5FS cross sections 
(``Santander matching"). The scheme is based on the observation that the 4FS and 5FS calculations 
of the cross section are better motivated in the asymptotic limits of small and large Higgs masses, 
respectively, and combines the two approaches in such a way that they are given variable weight, depending on 
the value of the Higgs-boson mass.

If both bottom jets accompanying the Higgs boson
in the final state are tagged, one has to rely on the fully exclusive
calculation for $gg\to b\bar b + h/H/A$. For the case of a single
$b$-tag in the final state the corresponding calculation in the 5FS
starts from the process $bg\to b+h/H/A$ with the final-state bottom
quark carrying finite transverse momentum. 
The NLO QCD, electroweak, and NLO SUSY-QCD corrections to this process have been calculated~\cite{bh5elw}.

State-of-the-art predictions as well as estimates of the corresponding
parametric and theoretical uncertainties, based on the described 
strategies and quoted calculations, have been provided by the 
LHC Higgs Cross Section Working Group both for 
total~\cite{Dittmaier:2011ti} and differential~\cite{Dittmaier:2012vm,Heinemeyer:2013tqa}
cross sections.

In this paper we are going to extend the previous 4FS calculations of
$t\bar t/b\bar b$+Higgs production cross sections at the LHC to include the full
supersymmetric QCD (SUSY-QCD) corrections within the MSSM
and separating the dominating universal part in terms of effective Yukawa
couplings. The paper is
organized as follows: In Section~\ref{se:calculation} we shall describe
the calculation of the NLO supersymmetric QCD corrections. Numerical
results for MSSM Higgs-boson production at the LHC are presented in
Section~\ref{se:results}. We conclude in Section~\ref{se:conclusions}.

\section{Calculational details}
\label{se:calculation}
\begin{figure}
\begin{picture}(100,500)(0,0)
\put(40.0,-10.0){\includegraphics{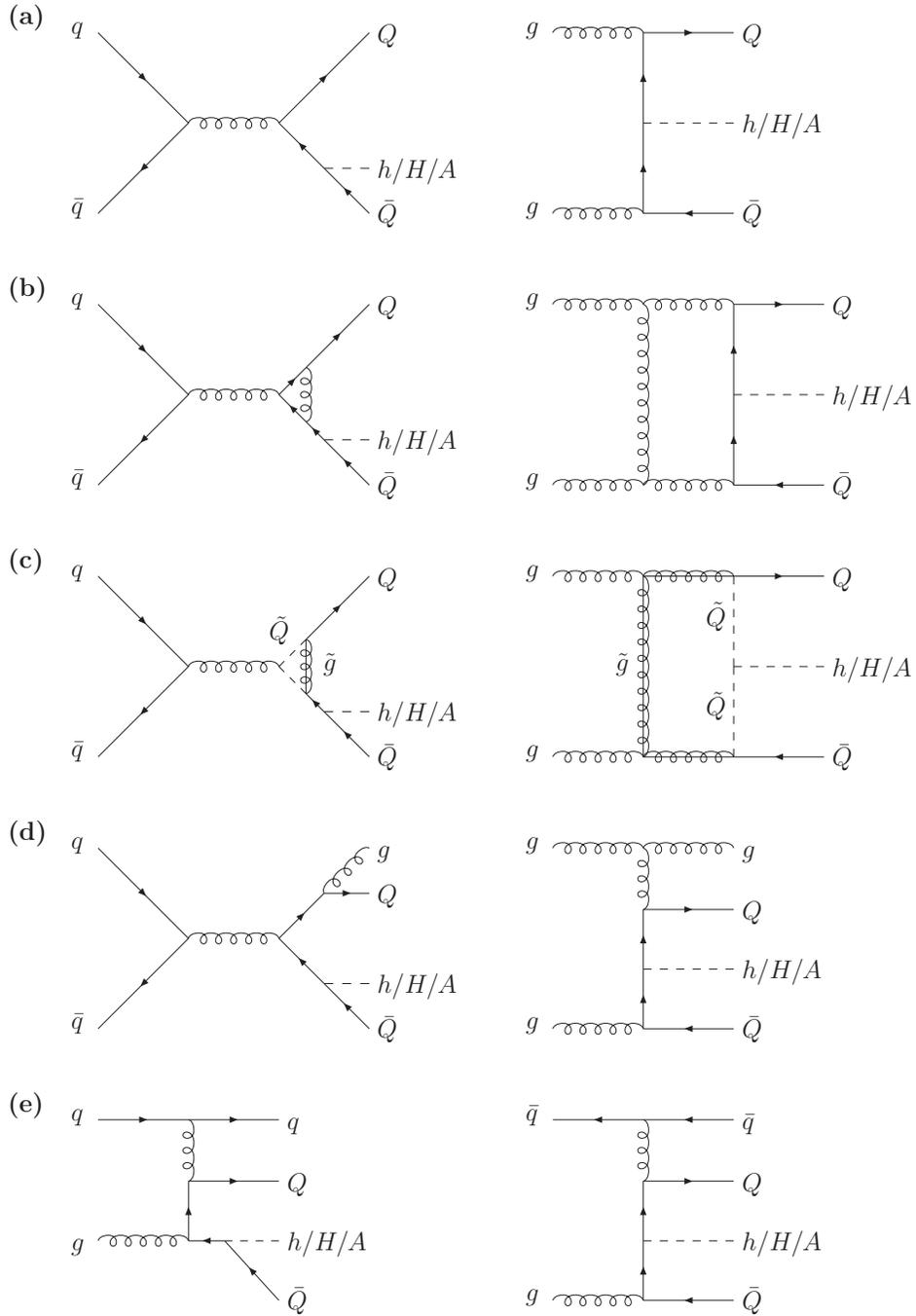}}
\end{picture}
\caption{A generic set of diagrams (a) for the Born level, (b) 
virtual gluon exchange, (c) virtual gluino and squark exchange, (d)
gluon radiation, and (e) gluon--(anti)quark scattering in the
subprocesses $q\bar q, gg\to Q\bar Q + h/H/A$, etc., where
$Q=t,b$.}
\label{fig:diags}
\end{figure}

At leading order (LO) Higgs radiation off heavy quarks is described by
the partonic processes $q\bar q, gg\to Q\bar Q+h/H/A$ ($Q=t,b$), as
depicted by the generic Feynman diagrams of Fig.~\ref{fig:diags}a.

The NLO corrections arise from virtual one-loop diagrams for the pure
QCD part (see Fig.~\ref{fig:diags}b), the genuine SUSY-QCD part (see
Fig.~\ref{fig:diags}c), the real corrections originating from gluon
radiation (see Fig.~\ref{fig:diags}d), and the crossed channel
emerging from gluon--quark initial states (see Fig.~\ref{fig:diags}e).
The details of our calculation
of the pure QCD corrections have been described in
Ref.~\cite{tthnlo} (see also Ref.~\cite{tbhnlo} for the closely related calculation 
of charged-Higgs-boson production with top and bottom quarks). Building on our previous papers, 
we have extended the calculations to include the genuine SUSY-QCD corrections.  
A more detailed
description of the corresponding computational details can be found in
Ref.~\cite{tthnlosusy}. Two independent calculations of the SUSY-QCD
corrections have been performed and found to be in mutual agreement.

The Feynman diagrams and amplitudes that contribute to the virtual
SUSY-QCD corrections have been generated with {\sl Feyn\-Arts}~3.2
\cite{Kublbeck:1990xc}. The amplitudes have been processed further with
two independent in-house {\sl Mathematica} routines, which automatically
create output in {\sl Fortran} and {\sl C++}, respectively. The genuine
virtual SUSY-QCD corrections are infrared finite, since all virtual
particles inside the loop contributions are massive. The pentagon tensor
integrals have been reduced directly to box integrals following
Ref.~\cite{Denner:2002ii}.  This method does not introduce inverse Gram
determinants in the reduction process, thereby avoiding numerical
instabilities in regions where these determinants become small. Box and
lower-point integrals have been reduced to scalar integrals using the
standard Passarino--Veltman technique~\cite{Passarino:1978jh}, 
Sufficient numerical stability is already achieved in this way, but further 
improvements could be achieved with the methods of Ref.~\cite{Denner:2005nn}
if needed in future evaluations.
The scalar integrals, finally, have been calculated either
analytically or using the results of Ref.~\cite{'tHooft:1978xw}. The
(IR-finite) scalar integrals have furthermore been checked with {\sl
LoopTools/FF}~\cite{Hahn:1998yk}.

The renormalization of the strong coupling $\alpha_{\mathrm{s}}(\mu)$
and the factorization of initial-state collinear singularities are
performed in the $\overline{\mathrm{MS}}$ scheme. The top quark and the
SUSY particles are decoupled from the running of
$\alpha_{\mathrm{s}}(\mu)$. In the 4FS calculation presented here for
the $b\bar b$+Higgs case, also the bottom quark is decoupled and the
partonic cross section is calculated using a four-flavour $\alpha_{\rm
s}$.  While the top- and bottom-quark masses are defined on-shell, the
$\overline{\mathrm{MS}}$ scheme is adopted for the renormalization of
the bottom--Higgs Yukawa coupling, which is fixed in terms of the
corresponding $\overline{\mathrm{MS}}$ renormalization of the bottom
mass with decoupled SUSY contributions. In order to sum large
logarithmic corrections $\propto \ln(m_b/\mu)$ we evaluate the Yukawa
coupling with the running $b$-quark mass
$\overline{m}_{b}(\mu)$~\cite{Braaten:1980yq}.

The SUSY loop corrections modify the tree-level relation between the
bottom mass and its Yukawa coupling, which is enhanced at large
$\tgb$~\cite{deltab}. These corrections can be summed to all orders
by replacing the bottom Yukawa coupling coefficients of
Table~\ref{tb:hcoup} by \cite{Carena:1999py,Guasch:2003cv}
\begin{eqnarray}
\tilde g_b^h & = & \frac{g_b^h}{1+\Delta_b} \left(
1-\frac{\Delta_b}{\tga~\tgb}\right), \nonumber \\
\tilde g_b^H & = & \frac{g_b^H}{1+\Delta_b} \left(
1+\Delta_b \frac{\tga}{\tgb}\right), \nonumber \\
\tilde g_b^A & = & \frac{g_b^A}{1+\Delta_b} \left(
1-\frac{\Delta_b}{\stgb}\right),
\label{eq:dmb}
\end{eqnarray}
where
\begin{eqnarray}
\Delta_b & = & \frac{C_F}{2}~\frac{\alpha_{\rm s}}{\pi}~m_{\sgl}~\mu~\tgb~
I(m^2_{\ssb_1},m^2_{\ssb_2},m^2_{\sgl}), \nonumber \\[2mm]
I(a,b,c) & = & \frac{\displaystyle ab\log(a/b) +
bc\log(b/c)
+ ca\log(c/a)}{(a-b)(b-c)(a-c)} 
\end{eqnarray}
with $C_F = 4/3$. Here, $\tilde{b}_{1,2}$ are the sbottom mass
eigenstates, and $m_{\tilde{g}}$ denotes the gluino mass.  The summation
formalism can be extended~\cite{Guasch:2003cv} to include corrections
proportional to the trilinear coupling $A_b$.  However, for the MSSM
scenarios under consideration in this work, these corrections turn out
to be small, and the corresponding summation effects may safely be
neglected. The scale of $\alpha_{\rm s}$ within the resummed bottom Yukawa
couplings of Eq.~(\ref{eq:dmb}) has been chosen as the average of the
contributing SUSY masses $\mu = (m_{\tilde{b}_1} + m_{\tilde{b}_2} +
m_{\tilde{g}})/3$, which is justified by the 
NNLO results for the
$\Delta_b$ corrections \cite{Noth:2008tw}. Note that the resummed bottom Yukawa
couplings have been defined by taking into account 5 active flavours. 
The strong coupling constant and the PDFs involved in the
explicit QCD and SUSY-QCD corrections to the production processes, on the other hand, 
are defined with four active flavours, analogous to the 4FS charged Higgs calculation discussed in 
some detail in Ref.~\cite{tbhnlo}.

If the LO cross section is expressed in terms of the bottom Yukawa
couplings including the summation of the $\tgb$-enhanced
corrections of Eq.~(\ref{eq:dmb}), the corresponding NLO contributions
have to be subtracted from the one-loop SUSY-QCD calculation to avoid
double counting. This subtraction is equivalent to an additional
finite renormalization of the bottom Yukawa coupling factors according to
\begin{equation}
g_b^\phi  \to  \tilde g_b^\phi \left[ 1 + \kappa_\phi \Delta_b \right] +
{\cal O}(\alpha_s^2)
\end{equation}
with
\begin{equation}
\kappa_h  =  1+\frac{1}{\tga~\tgb}, \quad
\kappa_H  =  1-\frac{\tga}{\tgb}, \quad
\quad {\rm and} \quad 
\kappa_A  =  1+\frac{1}{\stgb} \; .
\end{equation}

As we shall demonstrate in the numerical analysis presented in
Section~\ref{se:results}, the SUSY-QCD radiative corrections are indeed
sizable at large $\tgb$.  After absorbing the $\tgb$-enhanced terms
into effective Yukawa couplings,
however, the remaining one-loop SUSY-QCD corrections are small, below
the per-cent level, for scenarios with large SUSY-particle
masses.

\section{Numerical results}
\label{se:results}

To illustrate the impact of the NLO SUSY-QCD corrections we present 
numerical results for the LHC with CM energies
of 7 and 14~TeV. Note that further results of our calculation, including those 
for 8~TeV and a wider range of Higgs masses, will be available 
from the web pages of the LHC Higgs Cross Section Working Group \cite{HiggsXSWGweb}.
The cross sections for $t\bar t\phi^0$ are presented in
the 5FS, while for $b\bar b\phi^0$ production we have adopted the 4FS.
We have used the corresponding five- and four-flavour MSTW2008 PDFs \cite{mstw5,mstw4} 
with their respective LO strong coupling normalized to
$\alpha_{\rm s}(M_Z)=0.13939~(0.13355)$ and NLO coupling normalized to
$\alpha_{\rm s}(M_Z)=0.12018~(0.11490)$ in the 5FS (4FS). The on-shell top
mass has been adopted as $m_t=172.5$ GeV and the on-shell bottom mass as
$m_b=4.75$~GeV corresponding to a $\overline{\rm MS}$ mass
$\overline{m}_b(\overline{m}_b)=4.40$~GeV. The pole masses enter the
phase-space integration as well as the corresponding quark propagators
of the virtual and real matrix elements. The top pole mass has been used
in addition for the top Yukawa couplings, while the bottom Yukawa
couplings have been evaluated in terms of the $\overline{\rm MS}$ mass
as described above.  Our default choice for the renormalization and
factorization scales is $\mu_R=\mu_F=m_t + M_\phi/2$ for $t\bar t\phi^0$
production and $\mu_R=\mu_F=(2m_b + M_\phi)/4$ for the $b\bar b\phi^0$
case, see Refs.~\cite{tthnlo,bbhnlo,tbhnlo}.

We have chosen the Snowmass point SPS5 for Higgs radiation off top
quarks and SPS1b for the bottom quark case \cite{snowmass}. The MSSM
parameters of these two benchmark scenarios are given by%
\footnote{It
should be noted that the Snowmass point SPS5 is excluded by the recent
searches for squarks and gluinos by the ATLAS and CMS experiments \cite{lhc_susy} and
the Snowmass point SPS1b is close to the recent exclusion contours.
However, our results are only marginally affected qualitatively by
slightly larger squark and gluino masses in general and thus serve as a
guideline for scenarios beyond the present LHC reach, see e.g.\ Ref.~\cite{AbdusSalam:2011fc}.} 

\begin{displaymath}
\begin{array}{lllll}
\underline{\rm SPS5:} &
\tgb = 5, & 
\mu = 639.8 \gev, & 
A_t  =  -1671.4 \gev, & 
A_b = -905.6 \gev, \\[2mm]
 &
m_{\gl} = 710.3 \gev, &  
m_{\sq_L} =  535.2 \gev, &  
m_{\ssb_R} = 620.5 \gev, & 
m_{\st_R} = 360.5 \gev.
\end{array}
\end{displaymath}

\begin{displaymath}
\begin{array}{lllll}
\underline{\rm SPS1b:} &
\tgb   =  30, & 
\mu    =  495.6 \gev, & 
A_t    =  -729.3 \gev, & 
A_b    =  -987.4 \gev, \\[2mm]
 & 
m_{\gl}  =  916.1 \gev, & 
m_{\sq_L}  =  762.5 \gev, & 
m_{\ssb_R}  =  780.3 \gev, & 
m_{\st_R}  =  670.7 \gev.
\end{array}
\end{displaymath}

The pseudoscalar Higgs mass is left as a free parameter in both scenarios in order to
scan the Higgs mass ranges. The corresponding squark masses and
couplings that enter our calculation have been calculated by using the
tree-level relations. The Higgs masses and couplings have been
determined from $\tgb$ and the pseudoscalar mass $M_A$ by taking into
account higher-order corrections up to two loops in the effective
potential approach \cite{effpot} as included in the program {\sc Hdecay}
\cite{hdecay}. For the Higgs mass and coupling determination a
five-flavour $\alpha_{\rm s}$ has been used normalized to
$\alpha_{\rm s}(M_Z)=0.120$. The running bottom Yukawa couplings has
been determined with $\alpha_{\rm s}(M_Z)=0.120~(0.139)$ at NLO (LO)
with consistent NLO (LO) running.

\begin{figure}
\begin{picture}(100,500)(0,0)
\put(40.0,120.0){\includegraphics{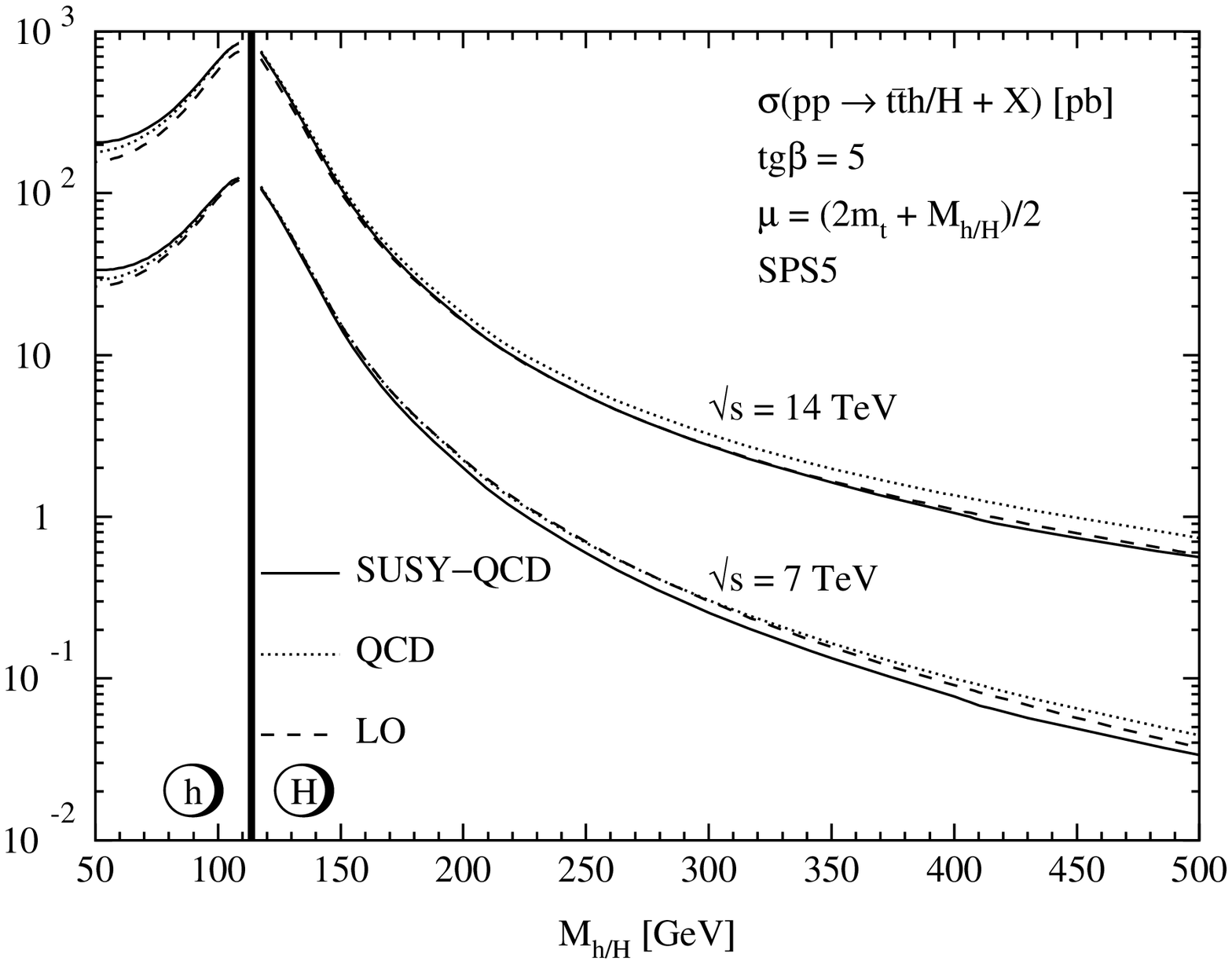}}
\put(40.0,-135.0){\includegraphics{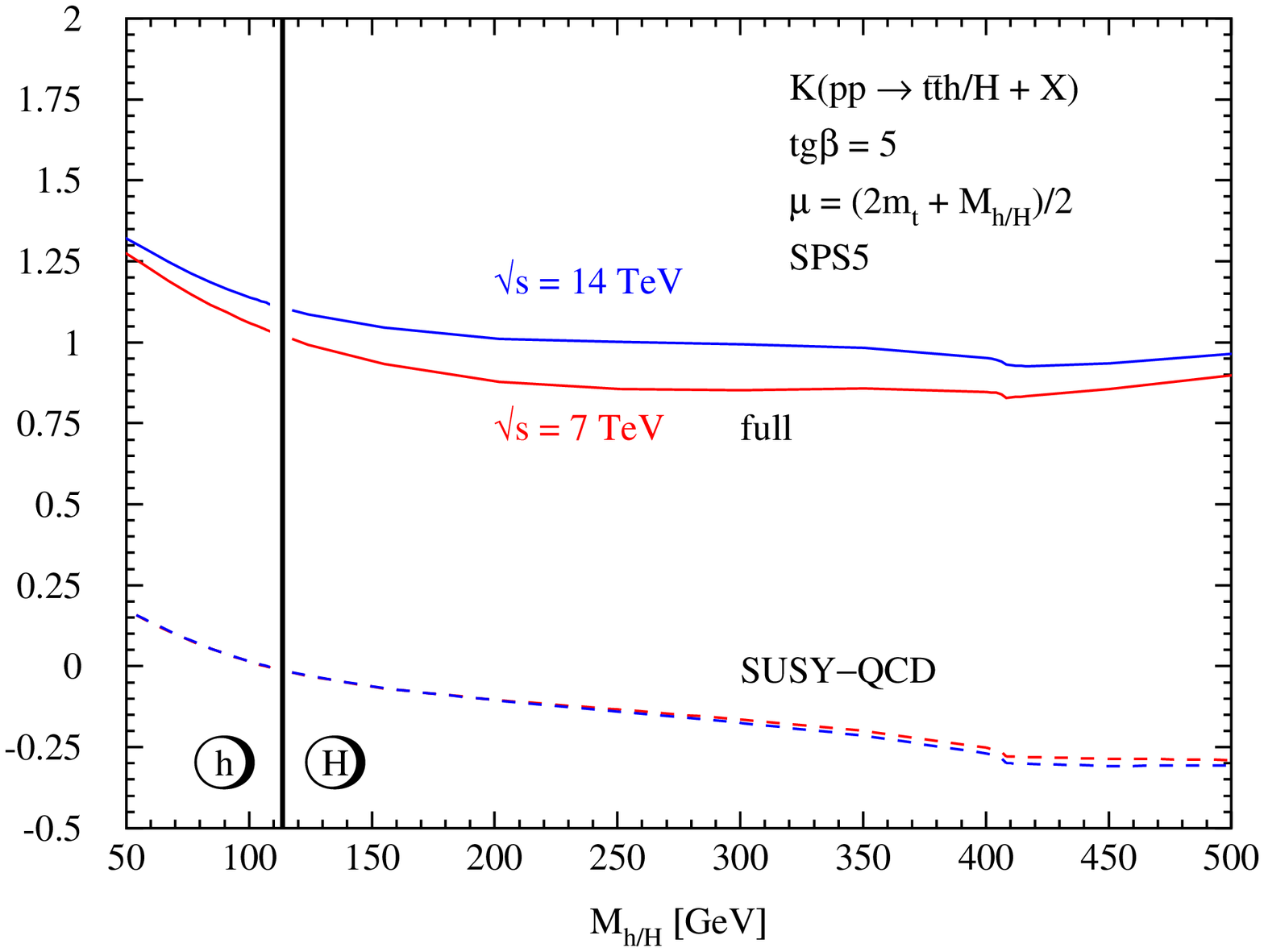}}
\end{picture}
\caption[]{\label{fg:cxn_tth} \it SUSY-QCD corrected production cross
sections of light and heavy scalar MSSM Higgs bosons in association with
$t\bar t$ pairs for the Snowmass point SPS5 \cite{snowmass}. {\it (a)}
Total cross sections, {\it (b)} $K$ factors.}
\end{figure}

The total cross sections for light and heavy scalar Higgs radiation off
top quarks are shown at LO and NLO in Fig.~\ref{fg:cxn_tth}a and the
corresponding $K$ factors in Fig.~\ref{fg:cxn_tth}b\footnote{We have
compared our calculation to the results for $t\bar t h$ production of
Ref.~\cite{tthnlosusy2}. Although we could not implement their scenario
exactly in our calculation we have found reasonable agreement within a
scenario very similar to theirs. The tiny SUSY--QCD corrections of
Ref.~\cite{tthnlosusy2} arise, since those authors work at the upper
bound of the light scalar Higgs mass where the genuine SUSY-QCD
corrections are small in our calculation, too, see
Fig.~\ref{fg:cxn_tth}b.}.
The contributions of the SUSY-QCD corrections to the $K$~factors, shown 
in the lower figure as dashed lines,
amount to about $\pm (20-30)\%$ and partially compensate the
pure QCD corrections for this MSSM scenario, thus leaving a moderate
total correction to the cross sections as can be inferred from
Fig.~\ref{fg:cxn_tth}b. The small kink at $M_H\sim 410$ GeV emerges from
the virtual $\tilde t_1 \overline{\tilde t}_1$ threshold%
\footnote{The
light stop mass amounts to $m_{\tilde t_1}=204.1$ GeV in the SPS5
scenario.} inside the virtual loop contributions.

\begin{figure}
\begin{picture}(100,500)(0,0)
\put(40.0,120.0){\includegraphics{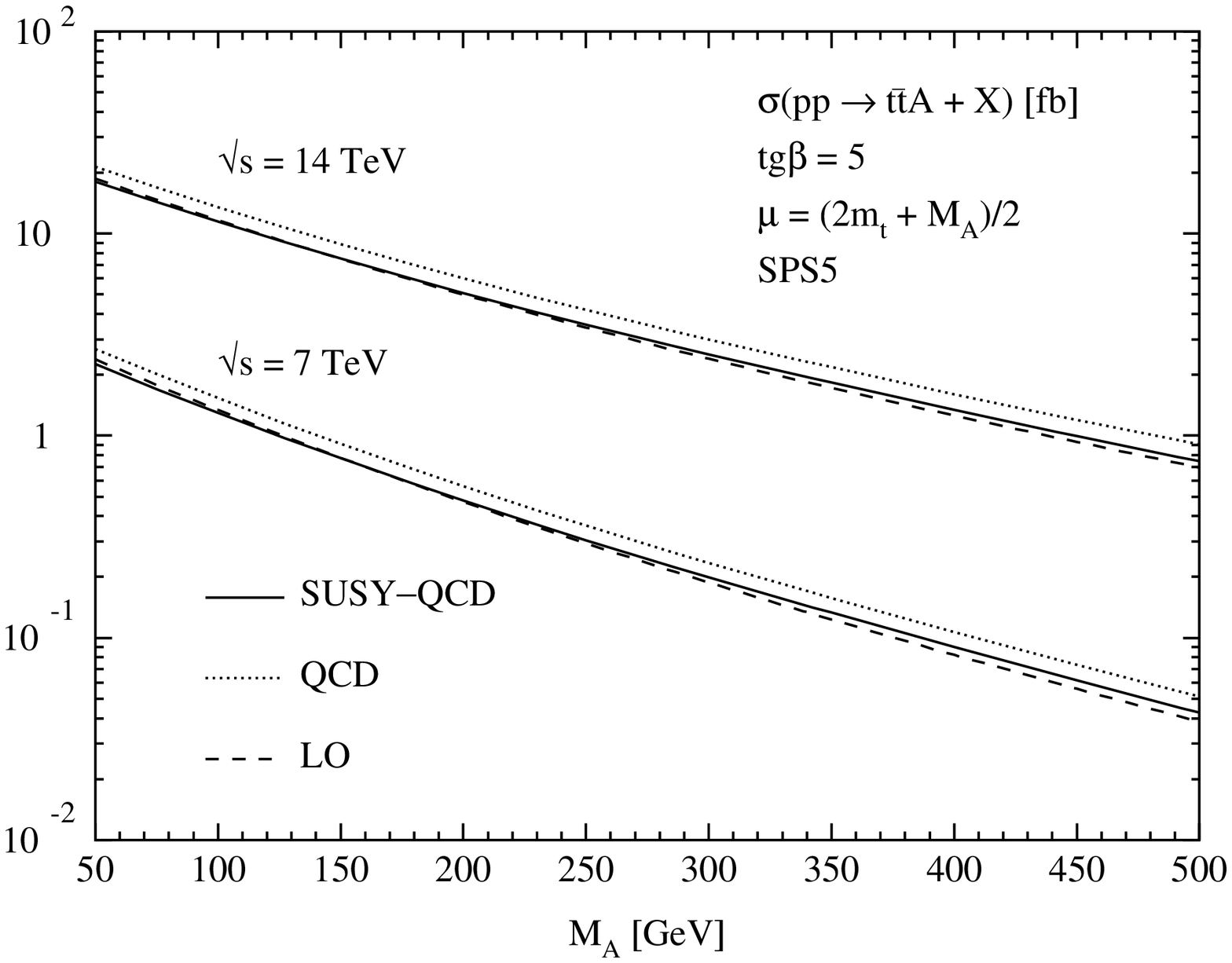}}
\put(40.0,-135.0){\includegraphics{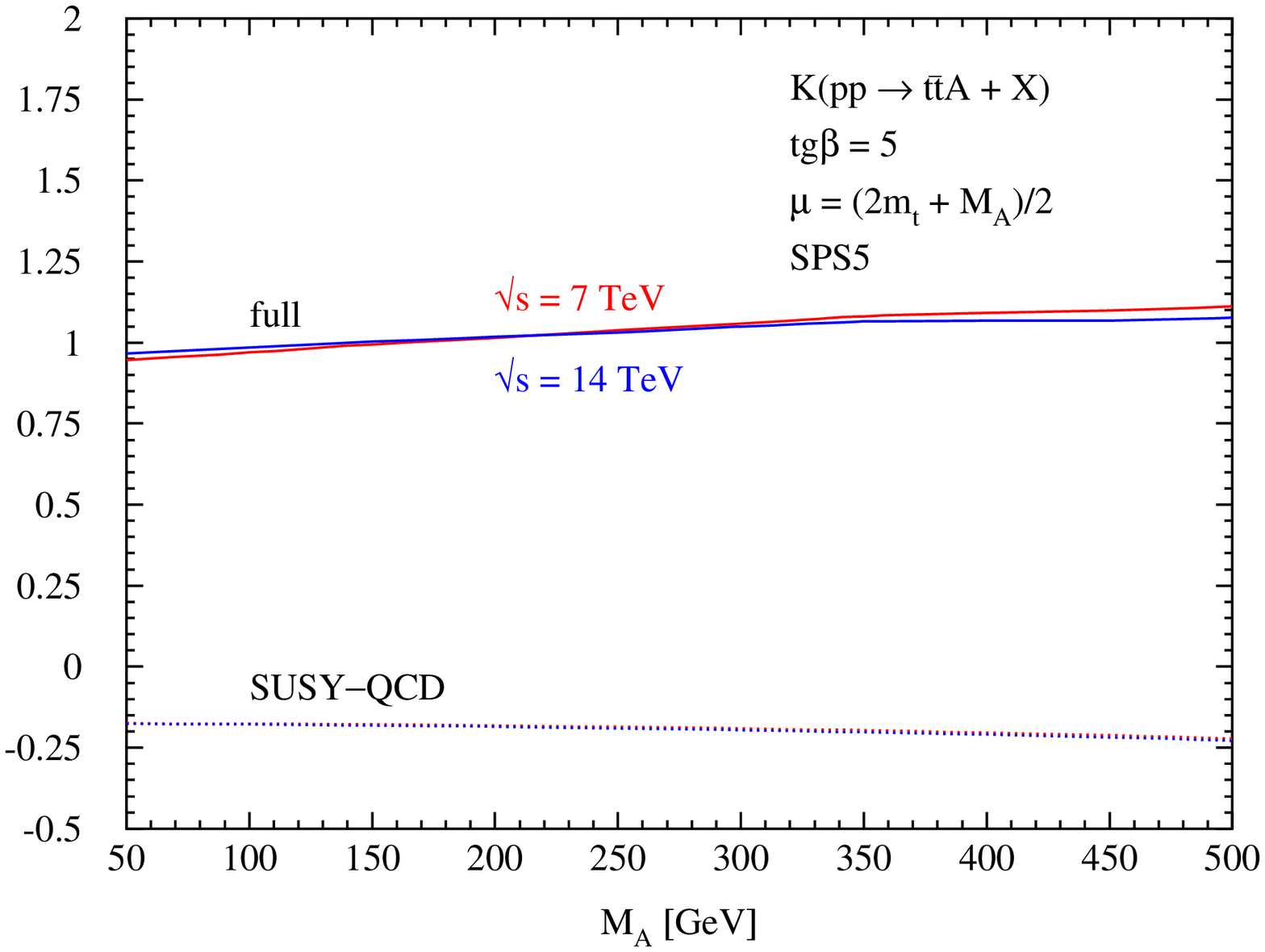}}
\end{picture}
\caption[]{\label{fg:cxn_tta} \it SUSY-QCD corrected production cross
sections of pseudoscalar MSSM Higgs bosons in association with $t\bar t$
pairs for the Snowmass point SPS5 \cite{snowmass}. {\it (a)} Total cross
sections, {\it (b)} $K$ factors.}
\end{figure}

The total cross sections for pseudoscalar Higgs radiation off top quarks
are displayed at LO and NLO in Fig.~\ref{fg:cxn_tta}a and the
corresponding $K$ factors in Fig.~\ref{fg:cxn_tta}b. 
Similar to the scalar
case the SUSY-QCD corrections amount to about --20\% and compensate the
pure QCD corrections for this MSSM scenario to a large extent, thus
leaving a moderate total correction to the cross sections as shown in
Fig.~\ref{fg:cxn_tta}b. The dependence of the $K$ factors on the LHC
CM energy is weaker than in the scalar Higgs case (cf.\ 
Fig.~\ref{fg:cxn_tth}b).

For Higgs-boson radiation off bottom quarks, we calculate the
NLO cross section according to
\beq
\sigma^\phi_{\rm NLO} = \sigma^\phi_0\times (1+\delta^\phi_{\rm SUSY})\times
(1+\delta^\phi_{\rm QCD}+\delta^\phi_{\rm SUSY-rem}),
\label{eq:corrbbh}
\eeq
where $\sigma^\phi_0$
denotes the LO cross section evaluated with LO $\alpha_{\rm s}$ and PDFs,
with the Yukawa coupling parametrized in terms
of the running b-quark mass $\overline{m}_b(\mu)$, but
without resummation of the $\tgb$-enhanced terms.
The correction $\delta^\phi_{\rm SUSY}$ comprises the $\tgb$-enhanced
terms according to Eq.~(\ref{eq:dmb}), including their resummation.
The remainder of the
genuine SUSY-QCD corrections is denoted by $\delta^\phi_{\rm SUSY-rem}$.

\begin{figure}
\begin{picture}(100,500)(0,0)
\put(40.0,120.0){\includegraphics{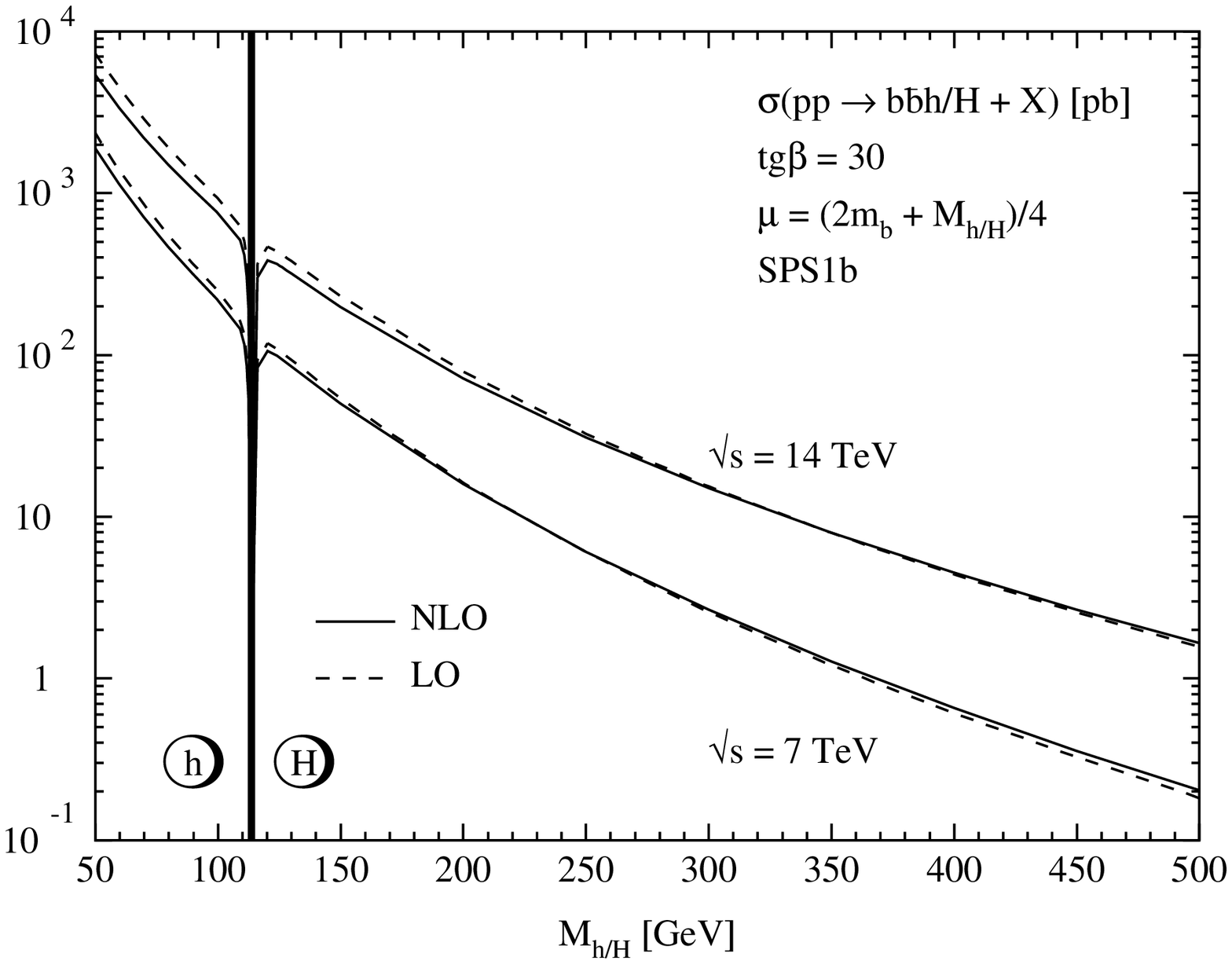}}
\put(40.0,-135.0){\includegraphics{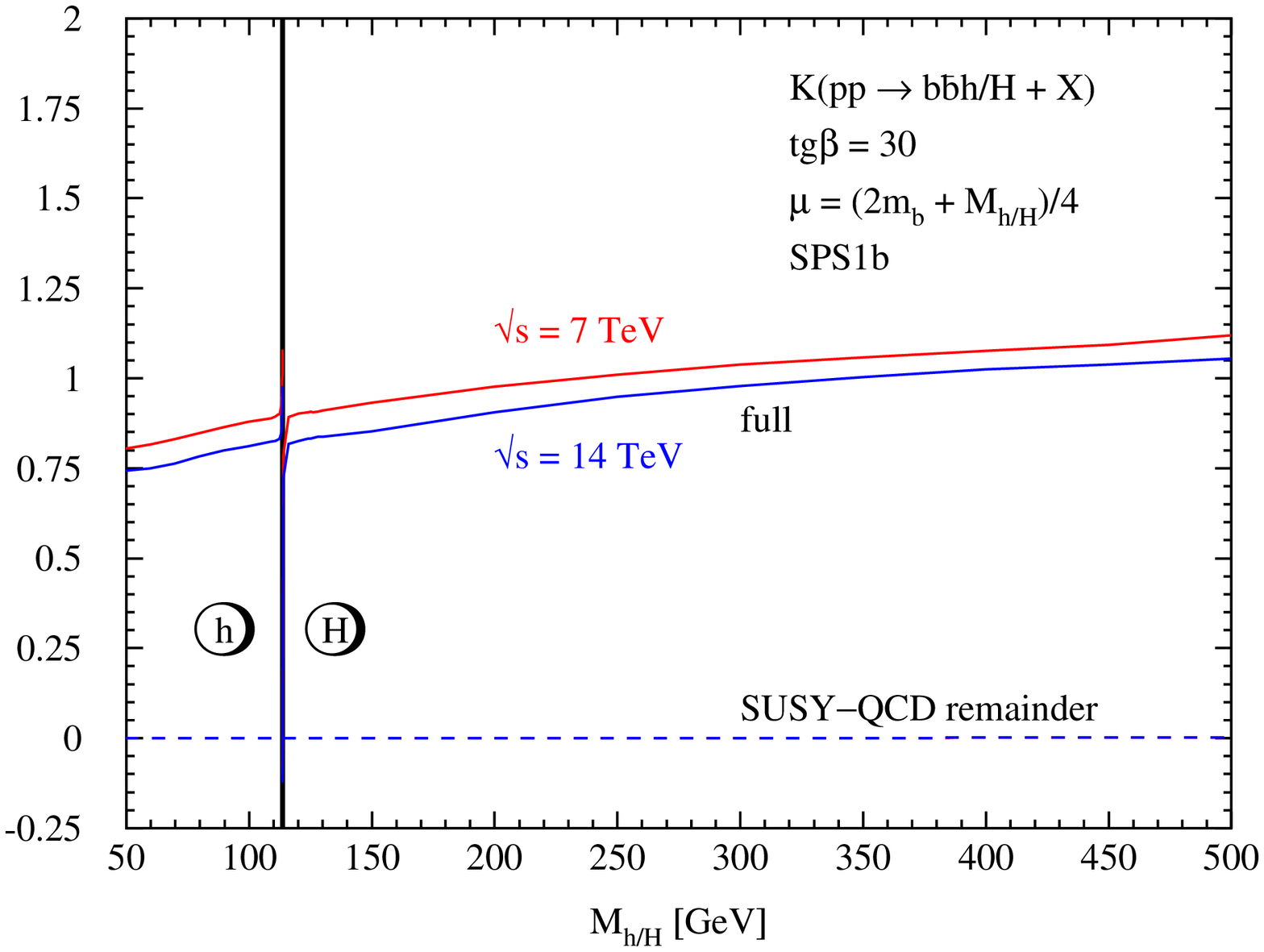}}
\end{picture}
\caption[]{\label{fg:cxn_bbh} \it SUSY-QCD corrected production cross
sections of light and heavy scalar MSSM Higgs bosons in association with
$b\bar b$ pairs for the Snowmass point SPS1b \cite{snowmass}. {\it (a)}
Total cross sections, {\it (b)} $K$ factors.}
\end{figure}

The results for scalar Higgs-boson radiation off bottom quarks are shown in
Fig.~\ref{fg:cxn_bbh}a (total cross section) and Fig.~\ref{fg:cxn_bbh}b
($K$ factor). 
Here we identify the LO cross section with $\sigma^\phi_0$,
which does not include the $\tgb$-enhanced resummation effects.
The observed moderate NLO corrections in this MSSM scenario, thus,
results from a compensation of the large QCD
corrections by large SUSY-QCD corrections.
The smallness of the SUSY-QCD remainder, however, shows that the
full NLO SUSY-QCD corrections are approximated extremely well by the 
$\tgb$-enhanced terms. 

Some numerical results for the  SUSY-QCD corrections to the process $pp \to b\bar{b} h$ 
have been presented in Ref.~\cite{Liu:2012qu}. However, only scatter plots are 
shown in ~\cite{Liu:2012qu}, but no
definite values for cross sections or SUSY-QCD corrections for a
reference input. It is thus not possible to quantitatively compare our
results to those of Ref.~\cite{Liu:2012qu}.

\begin{figure}
\begin{picture}(100,500)(0,0)
\put(40.0,120.0){\includegraphics{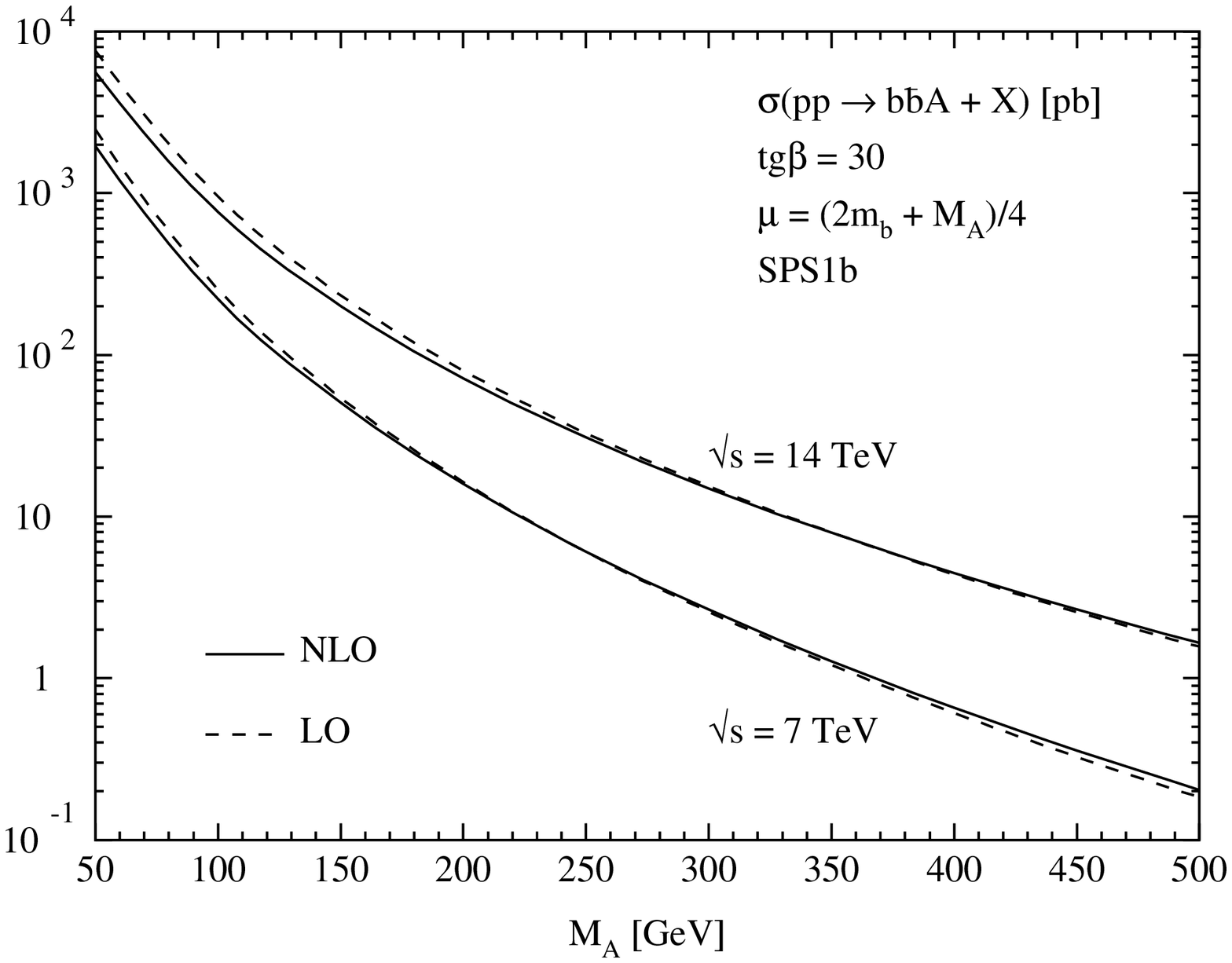}}
\put(40.0,-135.0){\includegraphics{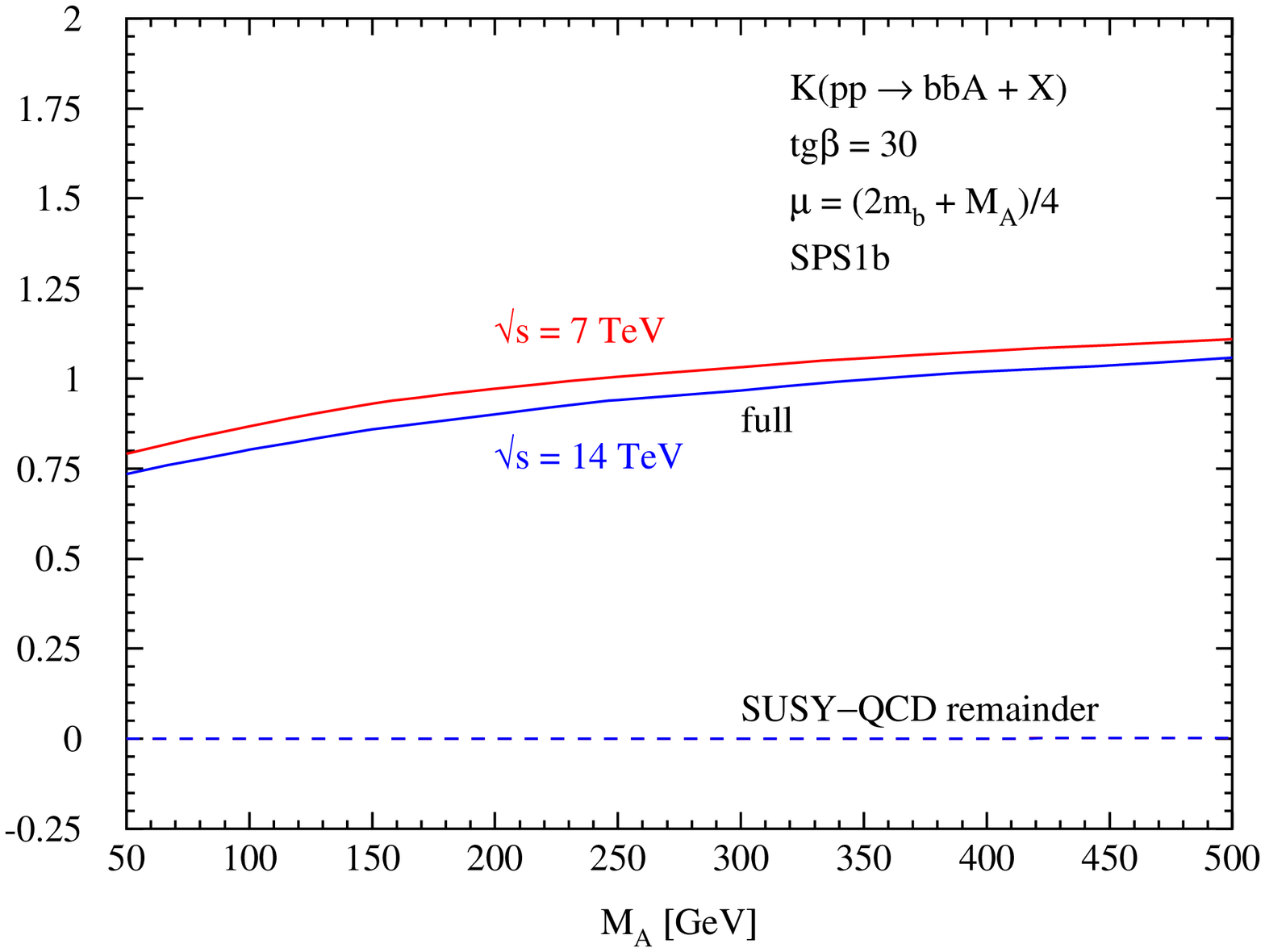}}
\end{picture}
\caption[]{\label{fg:cxn_bba} \it SUSY-QCD corrected production cross
sections of pseudoscalar MSSM Higgs bosons in association with $b\bar b$
pairs for the Snowmass point SPS1b \cite{snowmass}. {\it (a)} Total cross
sections, {\it (b)} $K$ factors.}
\end{figure}

Pseudoscalar Higgs radiation off
bottom quarks exhibits the same qualitative features as scalar Higgs production, as can be inferred
from Fig.~\ref{fg:cxn_bba}a for the total cross section and from 
Fig.~\ref{fg:cxn_bba}b for the $K$ factor. 
The CM energy dependence of the $K$ factor for pseudoscalar Higgs
production is similar to that of scalar Higgs production. The
pseudoscalar-Higgs-boson cross section at NLO coincides with the scalar
cross sections for the same Higgs mass values within a few per cent
apart from the regions where the light (heavy) scalar Higgs boson is
close to its upper (lower) mass bound.

In Table~\ref{tb:susy-qcd} we show the individual contributions to the
NLO cross sections for heavy scalar and pseudoscalar Higgs radiation off
bottom quarks. 
\begin{table}
\begin{center}
\begin{tabular}{|l||cl||ccc|ccc|} \hline
& $M_A$ & $M_H$ [GeV] & $\delta^A_{QCD}$ & $\delta^A_{SUSY}$ &
$\delta^A_{SUSYrem}$ & $\delta^H_{QCD}$ & $\delta^H_{SUSY}$ &
$\delta^H_{SUSYrem}$ \\
\hline \hline
       & 100 & 113.9 & 0.23 & $-0.30$ & $0.4\times 10^{-4}$  & 0.27 & $-0.38$ & $0.3\times 10^{-4}$ \\
       & 200 & 200   & 0.38 & $-0.30$ & $2.9\times 10^{-4}$  & 0.39 & $-0.30$ & $5.8\times 10^{-4}$  \\
7 TeV  & 300 & 300   & 0.46 & $-0.30$ & $6.7\times 10^{-4}$  & 0.47 & $-0.30$ & $9.3\times 10^{-4}$  \\
       & 400 & 400   & 0.53 & $-0.30$ & $1.3\times 10^{-3}$  & 0.53 & $-0.30$ & $1.5\times 10^{-3}$  \\
       & 500 & 500   & 0.57 & $-0.30$ & $2.0\times 10^{-3}$  & 0.59 & $-0.30$ & $2.2\times 10^{-3}$  \\
\hline
       & 100 & 113.9 & 0.14 & $-0.30$ & $0.4\times 10^{-4}$  & 0.17 & $-0.38$ & $0.5\times 10^{-4}$ \\
       & 200 & 200   & 0.28 & $-0.30$ & $2.7\times 10^{-4}$  & 0.29 & $-0.30$ & $5.7\times 10^{-4}$  \\
14 TeV & 300 & 300   & 0.37 & $-0.30$ & $6.5\times 10^{-4}$  & 0.39 & $-0.30$ & $9.3\times 10^{-4}$  \\
       & 400 & 400   & 0.45 & $-0.30$ & $1.2\times 10^{-3}$  & 0.45 & $-0.30$ & $1.5\times 10^{-3}$  \\
       & 500 & 500   & 0.50 & $-0.30$ & $2.1\times 10^{-3}$  & 0.49 & $-0.30$ & $2.3\times 10^{-3}$  \\
\hline
\end{tabular}
\caption[]{\label{tb:susy-qcd} \it Individual NLO corrections relative
to the LO cross section for $pp\to b\bar b \phi + X$ ($\phi=H,A$),
as defined in Eq.~(\ref{eq:dmb}),
are shown for the
LHC at two CM energies (7~TeV and 14~TeV) for the Snowmass point SPS1b
\cite{snowmass}.}
\end{center}
\vspace*{1em}
\begin{center}
\begin{tabular}{|l|c||cl||cc|cc|} \hline
& $\tgb$ & $M_A$ & $M_H$ [GeV] & $\delta^A_{SUSY}$ & $\delta^A_{SUSYrem}$
& $\delta^H_{SUSY}$ & $\delta^H_{SUSYrem}$ \\
\hline \hline
      & 3  & 200 & 209.7 & $-0.04$ & $2.1\times 10^{-4}$ & $-0.04$ & $5.7\times 10^{-4}$ \\
      & 5  & 200 & 204.0 & $-0.06$ & $2.4\times 10^{-4}$ & $-0.06$ & $5.3\times 10^{-4}$  \\
      & 7  & 200 & 202.1 & $-0.08$ & $2.5\times 10^{-4}$ & $-0.09$ & $3.9\times 10^{-4}$  \\
7 TeV & 10 & 200 & 200.9 & $-0.12$ & $2.5\times 10^{-4}$ & $-0.12$ & $3.8\times 10^{-4}$  \\
      & 20 & 200 & 200.1 & $-0.21$ & $2.6\times 10^{-4}$ & $-0.21$ & $4.4\times 10^{-4}$  \\
      & 30 & 200 & 200.0 & $-0.30$ & $2.9\times 10^{-4}$ & $-0.30$ & $5.8\times 10^{-4}$  \\
\hline
       & 3  & 200 & 209.7 & $-0.04$ & $2.0\times 10^{-4}$ & $-0.04$ & $7.2\times 10^{-4}$ \\
       & 5  & 200 & 204.0 & $-0.06$ & $2.2\times 10^{-4}$ & $-0.06$ & $5.0\times 10^{-4}$  \\
       & 7  & 200 & 202.1 & $-0.08$ & $2.4\times 10^{-4}$ & $-0.09$ & $4.4\times 10^{-4}$  \\
14 TeV & 10 & 200 & 200.9 & $-0.12$ & $2.5\times 10^{-4}$ & $-0.12$ & $4.1\times 10^{-4}$  \\
       & 20 & 200 & 200.1 & $-0.21$ & $2.7\times 10^{-4}$ & $-0.21$ & $4.4\times 10^{-4}$  \\
       & 30 & 200 & 200.0 & $-0.30$ & $2.7\times 10^{-4}$ & $-0.30$ & $5.7\times 10^{-4}$  \\
\hline
\end{tabular}
\caption[]{\label{tb:tgb-dep} \it The $\tgb$ dependence of the
individual genuine SUSY-QCD corrections relative to the LO cross
section for $pp\to b\bar b \phi + X$ ($\phi=H,A$) at the LHC for two
CM energies (7~TeV and 14~TeV). 
The Snowmass point SPS1b \cite{snowmass} is
adopted with $M_A$ and $\tgb$ left as free parameters. The
pseudoscalar Higgs mass is fixed at 200~GeV.}
\end{center}
\end{table}
The relative pure QCD corrections $\delta^\phi_{\rm QCD}$
($\phi=H,A$), the $\tgb$-enhanced SUSY-QCD corrections $\delta^\phi_{\rm SUSY}$,
and the remainders of the SUSY-QCD corrections
$\delta^\phi_{\rm SUSY-rem}$ are defined according to Eq.~(\ref{eq:corrbbh}).
As already observed in the corresponding figures,
the sizable QCD corrections $\delta^\phi_{\rm QCD}$ are to a large extent compensated by the
SUSY-QCD corrections $\delta^\phi_{\rm SUSY}$, thus
leaving a small remainder $\delta^\phi_{\rm SUSY-rem}$ of the SUSY-QCD
corrections below the per-cent level for all Higgs masses in the SPS1b
scenario.

In order to quantify the accuracy of the $\Delta_b$ approximation of
Eq.~(\ref{eq:dmb}) we display the effect of the $\tgb$ resummation and
the corresponding remainders of the SUSY-QCD corrections for various
values of $\tgb$ in Table~\ref{tb:tgb-dep}.  For these numbers we have
chosen the Snowmass point SPS1b, but set $M_A=200$ GeV and vary 
$\tgb$ between $3 \le \tgb \le 30$. The results demonstrate that
the $\Delta_b$ approximation for the genuine SUSY-QCD corrections is
accurate to better than the per-cent level for all values of $\tgb$, at least in MSSM scenarios 
with heavy SUSY particles. 
Similar results have been obtained for other Higgs masses within the
SPS1b scenario.

\section{Conclusions}
\label{se:conclusions}

We have presented the next-to-leading order supersymmetric QCD
corrections to neutral MSSM Higgs-boson production at the LHC 
through the parton processes $q\bar q,gg \to Q\bar Q + h/H/A$ ($Q=t,b$).
For neutral Higgs radiation off top quarks we find moderate SUSY-QCD corrections, 
similar in magnitude to the pure QCD effects. Depending on the MSSM 
scenario, the genuine SUSY-QCD corrections 
can be positive or negative. On the other hand, the genuine
SUSY-QCD corrections to neutral Higgs radiation off bottom quarks turn
out to be large for large values of $\tgb$, due to large $\tgb$-enhanced
contributions. They can thus significantly change the large pure QCD
corrections. The dominant part of these genuine
SUSY-QCD corrections can be absorbed in effective bottom Yukawa
couplings which resum the $\tgb$-enhanced corrections. Although not
explicitly shown in this work the same conclusions hold for the
differential cross sections. The NLO results obtained in this work
will be important for a reliable extraction of the SUSY parameters from
future LHC data if the new particle with mass $\approx 125$ GeV turns out
to be a supersymmetric Higgs boson.\\

\noindent
{\bf Acknowledgements.}
This work has been supported by the DFG SFB/TR9 ``Computational Particle Physics''.
The research of S.D.\ and M.S.\ is supported in part by the European 
Commission through the ``HiggsTools'' Initial Training Network 
PITN-GA-2012-316704.

\end{document}